\documentclass[runningheads]{llncs}
\usepackage[T1]{fontenc}
\usepackage{graphicx}
\usepackage{orcidlink}
\usepackage{amsmath}
\usepackage{hyperref}
\begin{document}
\title{Self-Efficacy and Favorability Shape Learning from Tutoring Systems and Paper Practice}
\titlerunning{Self-Efficacy and Favorability Shape Learning}
\author{
Xinfei Cen\inst{1}\orcidlink{0009-0007-9555-7010}\thanks{Corresponding author.} \and
Vincent Aleven\inst{1}\orcidlink{0000-0002-1581-6657} \and
Kenneth R. Koedinger\inst{1}\orcidlink{0000-0002-5850-4768} \and
Conrad Borchers\inst{2}\orcidlink{0000-0003-3437-8979} \and
Paulo F. Carvalho\inst{1}\orcidlink{0000-0002-0449-3733}
}

\authorrunning{X. Cen et al.}

\institute{
Carnegie Mellon University, Pittsburgh, PA, USA \\
\email{\{xinfeic,aleven,pcarvalh\}@cs.cmu.edu, koedinger@cmu.edu}
\and
Vanderbilt University, Nashville, TN, USA\\
\email{c.borchers@vanderbilt.edu}
}

\maketitle              %
\begin{abstract}
Motivational factors such as self-efficacy and how favorably students feel toward practice play a crucial role in shaping learning, particularly in technology-supported environments. Yet, educational interventions often overlook how these factors interact with practice format. This paper examines the influence of self-efficacy and favorability on learning outcomes across two common practice formats: paper-based and system-based tutoring practice. Using a counterbalanced within-subject design with matched problem sets, we isolate the effect of practice format while modeling motivational differences. Results indicate that students with lower baseline self-efficacy achieved greater learning gains regardless of practice format. Among students with lower baseline self-efficacy, greater favorability toward the tutor was associated with greater learning gains during tutor practice, whereas the pattern differed in paper-based practice. Intelligent Tutoring System (ITS)-based practice did not significantly improve post-training self-efficacy relative to paper-based methods. These findings underscore the potential value of tailoring practice format to students’ motivational profiles, as the benefits of tutor- and paper-based practice varied with baseline self-efficacy and favorability. They lay the groundwork for future research on how instructional formats can be aligned more effectively with learners’ motivational needs.

\end{abstract}

\section{Introduction}

Technology-enhanced learning (TEL) systems that support learning by problem solving are widely used in domains such as mathematics and science. These systems can improve learning outcomes through features such as scaffolded support, step-level feedback, and individualized practice~\cite{Ma:2014,VanLehn:2011}. Beyond learning performance, however, TEL systems may also shape learners’ motivational experiences during practice. Likewise, learners’ motivational profiles, including self-efficacy and favorability toward different instructional formats, may influence how they respond to those formats and whether they benefit differently from one format than another. Prior work has shown that motivational factors such as self-efficacy play an important role in students' engagement and learning~\cite{Bandura:1997}, and that learners’ confidence and comfort with digital tools can support motivation and engagement in technology-based learning~\cite{Pan:2020}. However, it remains less clear how motivational factors interact with instructional format in content-matched tutor- and paper-based practice, and whether these formats differentially influence students’ self-efficacy. Understanding these relationships is therefore important for designing and using TEL systems in ways that better support diverse learners. In this paper, we examine both how learners’ motivational profiles shape learning outcomes in tutor- and paper-based practice and whether these instructional formats, in turn, differentially influence students’ self-efficacy.

\section{Related Work}

Self-efficacy, students’ belief in their ability to succeed on a task, is a well-established predictor of academic outcomes, especially in self-directed learning contexts where students must actively manage their own learning. High self-efficacy supports engagement, persistence, and resilience in the face of setbacks~\cite{Bandura:1997,Cook:2016,Zimmerman:2000}. Research has shown that self-efficacy can be shaped over time by multiple factors, such as instruction, feedback, or students’ perceived success~\cite{Usher:2008}. Favorability, or students’ attitudes or preferences toward instruction, also plays an important motivational role. Research shows that when students are more favorable toward learning materials and find them personally meaningful or interesting, they are more likely to engage deeply and persist, which can energize learning~\cite{Harackiewicz:2016}. Both self-efficacy and favorability are critical in technology-supported learning environments. Learners’ confidence and comfort with digital tools can positively shape their motivation and engagement in technology-based learning, which in turn supports learning outcomes~\cite{Pan:2020}.

Intelligent Tutoring Systems (ITSs) provide scaffolded support, immediate feedback, and individualized pacing—features that have consistently improved learning outcomes in domains such as math and science ~\cite{Anderson:1995,Kulik:2016,Ma:2014,VanLehn:2011}. ITSs can also support metacognition through self-explanation, adaptive strategies to reduce gaming behaviors, and enhanced help-seeking skills~\cite{Aleven:2002,Baker:2006}. These supports may also be relevant to learners’ self-efficacy, as perceived success and performance feedback offered by ITSs can contribute to mastery experiences and social persuasion, two well-established sources of self-efficacy beliefs~\cite{Usher:2008,Bandura:1997}. In contrast, paper-based practice relies more heavily on students’ independent effort, placing greater demands on metacognitive and self-regulatory skills. Although ITSs have often been studied in relation to conventional classroom instruction, where paper-based practice remains widely used~\cite{Gray:2020}, these comparisons are often confounded by differences in curriculum, pedagogy, or task type~\cite{VanLehn:2011}.

Despite decades of ITS research, three major gaps remain. First, as mentioned before, with the exception of a small number of content-matched studies~\cite{Mendicino:2009}, most studies do not control for instructional content when comparing ITS and paper-based formats, making it difficult to isolate the effect of the practice condition itself. Second, while self-efficacy and favorability are known to influence engagement and academic performance~\cite{Schunk:2009,Cordova:1996,Harackiewicz:2016}, few studies have systematically examined how these motivational factors interact with instructional format to shape learning outcomes. Third, given their scaffolded support and immediate feedback, ITSs may plausibly foster self-efficacy through perceived success and mastery experiences~\cite{VanLehn:2011,Bandura:1997}. Some recent work has observed self-efficacy gains after tutoring-system practice~\cite{Conrad:2025}. However, relatively few studies have directly tested this claim, and the available evidence is limited~\cite{Bernacki:2014,McQuiggan:2007}.

Prior work directly compared the benefits of paper-based practice and tutor-based practice, a form of ITS, for learning outcomes~\cite{Borchers:2023,Borchers:2023Correction}. However, questions remain regarding the role of students' self-efficacy and preferences in such comparisons. Tutor-based practice provides scaffolded questions and timely feedback that may especially benefit students with high baseline self-efficacy and further promote their self-efficacy, whereas paper-based practice relies more on learner autonomy. In the present study, we analyze questionnaire data collected in prior work~\cite{Borchers:2023} to investigate two research questions: 
\begin{enumerate}
    \item RQ1: Do baseline self-efficacy and favorability influence learning outcomes in tutor- versus paper-based practice?
    \item RQ2: Do ITSs promote self-efficacy compared to paper-based practice?
\end{enumerate}

As emphasized by recent work, it is critical to move beyond general instructional claims~\cite{Yan:2023}, examine how learner-specific motivational factors interact with different practice formats, and work toward identifying best practices that support all learners more effectively.

\section{Methods}

\subsection{Sample and Experimental Design}

\begin{figure}
  \centering
  \includegraphics[
  width=0.8\textwidth
  ]{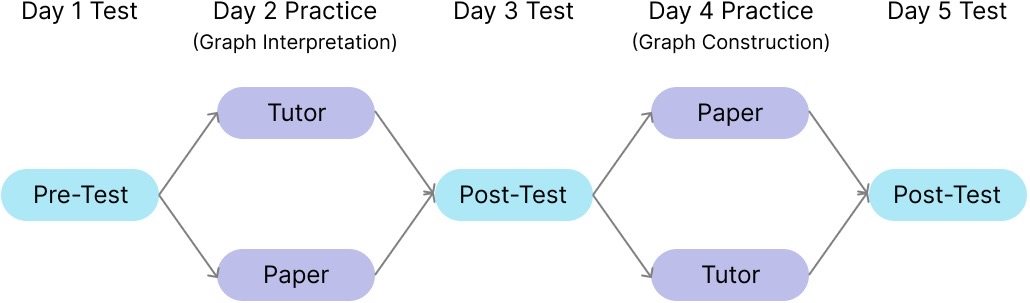}
  \caption{Within-subjects crossover experiment design.}
  \label{fig:experiment}
\end{figure}

\begin{figure}
  \centering
  \includegraphics[
  width=0.8\textwidth
  ]{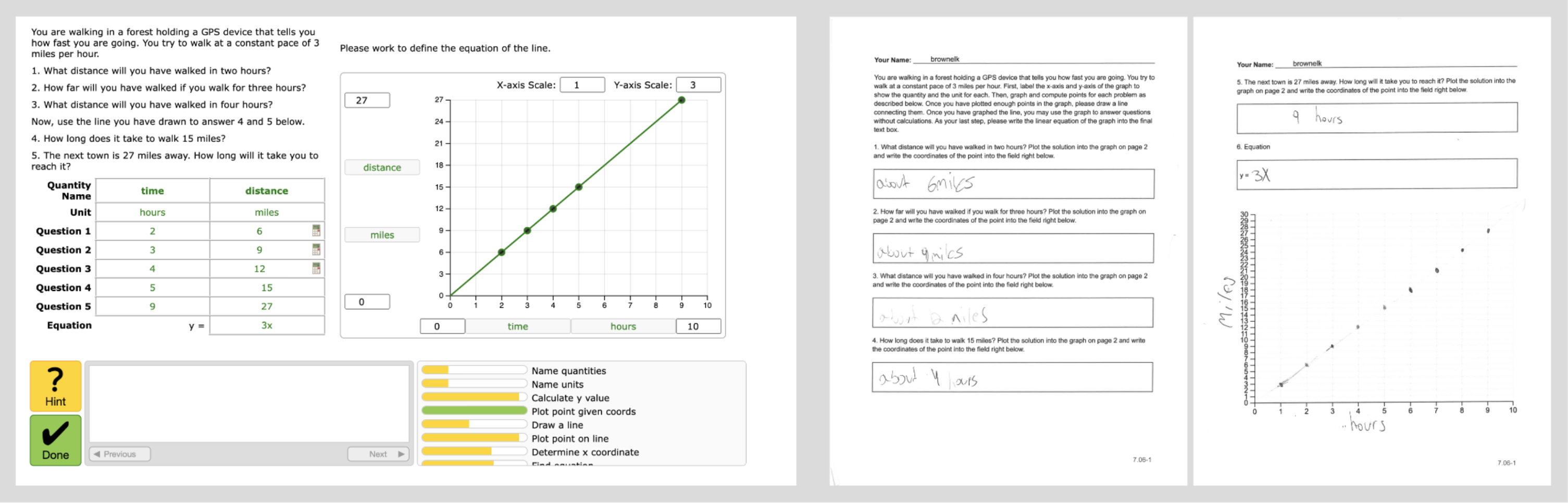}
  \caption{Tutor-based interface (left) and paper-based materials (right) with example student responses from a graph construction problem.}
  \label{fig:interface}
\end{figure}

The dataset includes 97 students from seven 9th-grade mathematics classes at a suburban middle school in the northeastern United States who completed pretest and posttest assessments on linear algebra–related topics and were instructed by three teachers. The original classroom study was conducted under an approved IRB protocol, and all data analyzed here were de-identified. There were two practice units in total (see Fig.~\ref{fig:experiment}): graph interpretation, practiced on Day 2, and graph construction, practiced on Day 4, with each student completing two approximately 40-minute practice sessions. These units focused, respectively, on interpreting linear graphs (e.g., reading values and relationships from graphs) and constructing linear graphs (e.g., plotting points, inferring values, and deriving linear equations). All students experienced both practice conditions (i.e., paper-based and tutor-based), but the order varied: each student received one unit using the tutor-based method and the other using the paper-based method. While the topic order was fixed, practice formats were counterbalanced across students, such that each format appeared in both the first and second practice positions. Although we cannot rule out the possibility of carryover effects, by which practice on the first topic may have improved performance on the second, this design reduced the likelihood that these carryover effects differed between the practice conditions. As illustrated in Fig.~\ref{fig:interface}, students in the tutor-based condition solved problems using an intelligent tutoring system that provided step-by-step guidance, including correctness feedback and hints. The paper-based problems served as a business-as-usual control but were designed to closely mirror the structure and problem-solving steps of the tutor-based problems, while omitting scaffolded support such as step-level feedback and hints. More details about the instructional materials, including examples of the paper-based practice materials, and study procedures can be found in~\cite{Borchers:2023}. Additionally, all students completed a pretest at the start of the study and a posttest at the end of each practice condition. The current work builds on prior analyses of the same data that focused only on learning outcomes~\cite{Borchers:2023} by examining motivational measures.

\subsection{Motivational Survey Instruments}

Along with the training and testing data, we collected students' self-efficacy responses before the pretest (baseline) and after the second posttest. The same questionnaires were used at both pretest and posttest. The self-efficacy questionnaire included six questions constructed following Bandura's guidelines for measuring self-efficacy~\cite{Bandura:2006}. All were 0--100 Likert-scale questions ranging from ``Cannot do at all'' to ``Highly certain I can do''. The first three questions were designed to assess students' self-efficacy in skills related to the first training unit, focusing on students’ ability to interpret and compare graph elements (e.g., ``Compare the slope and intercept of a graph''). The other three questions evaluated students' confidence in skills related to the second training unit, including plotting points, labeling axes, and generalizing concepts to real-world situations (e.g., ``Decide what numbers to place along each axis''). Additionally, the favorability questions assessed students' attitudes toward the two instructional methods (i.e., paper-based and tutor-based practice; e.g., ``I like doing math problems with the computer tutor'') using a Likert scale (1: strongly agree - 5: strongly disagree), with one additional free-response question for further feedback.\footnote{The full questionnaires are available on OSF: \url{https://osf.io/r5e9z/}.} 

All data were collected using paper-based surveys, and a significant proportion of students left at least one question blank. We conducted factor analysis and reliability testing to identify meaningful constructs. Specifically, two self-efficacy constructs (graph interpretation and graph construction) and two favorability constructs (tutor-based and paper-based) were extracted. Construct scores were computed as the average of the items associated with each construct, using available responses when individual items were missing. Students were included in the final analyses if construct scores could be computed and complete performance data were available for both practice conditions, resulting in a final analytic sample of 54 students. All constructs demonstrated adequate internal consistency (Cronbach’s $\alpha$): graph interpretation ($\alpha_{\text{pre}}=.83$, $\alpha_{\text{post}}=.83$), graph construction ($\alpha_{\text{pre}}=.81$, $\alpha_{\text{post}}=.73$), tutor-based favorability ($\alpha=.72$), and paper-based favorability ($\alpha=.71$).

\subsection{Modeling}

To investigate \textbf{RQ1}—whether baseline self-efficacy and favorability influence learning outcomes in tutor- versus paper-based practice—we conducted a linear mixed-effects analysis. Our model included students' learning gain (pretest-to-posttest performance) as the dependent variable, with practice condition (tutor-based vs. paper-based training), baseline self-efficacy, and favorability toward the respective practice condition as fixed effects. Additionally, we included a three-way interaction among condition, baseline self-efficacy, and favorability to examine their combined influence on student learning outcomes. All changes in performance scores, baseline self-efficacy, post-training self-efficacy, and favorability were normalized to estimate effect sizes, and student ID was modeled as a random effect to account for repeated tests. This approach allowed us to assess how baseline self-efficacy, practice condition, and favorability interact to shape students' learning outcomes. To examine \textbf{RQ2}—whether ITSs promote self-efficacy compared to paper-based practice—we used the same linear mixed-effects model as before, replacing learning gain with post-training self-efficacy as the dependent variable.

\section{Results}

\subsection{RQ1: How Baseline Self-Efficacy and Favorability Shape Learning}

We first examined whether baseline self-efficacy and favorability influenced learning gains across training conditions (tutor-based vs. paper-based) using a linear mixed-effects model (see Table~\ref{tab:rq1}).

\begin{table}
\centering
\caption{Fixed-effects model results for RQ1.}
\label{tab:rq1}
\resizebox{\textwidth}{!}{
\begin{tabular}{lccccc}
\hline
Effect & Estimate & Std. Error & df & $t$ value & $p$ \\
\hline
Intercept & 0.085 & 0.117 & 64.179 & 0.728 & 0.469 \\
Practice Format (Paper vs. Tutor) & 0.532 & 0.223 & 62.303 & 2.383 & 0.020 \\
Baseline Self-Efficacy & -0.283 & 0.132 & 98.403 & -2.136 & 0.035 \\
Favorability & 0.078 & 0.114 & 91.120 & 0.686 & 0.494 \\
Practice Format × Baseline Self-Efficacy & -0.110 & 0.266 & 100.000 & -0.412 & 0.681 \\
Practice Format × Favorability & 0.174 & 0.232 & 96.796 & 0.750 & 0.455 \\
Baseline Self-Efficacy × Favorability & -0.044 & 0.122 & 99.435 & -0.357 & 0.722 \\
Practice Format × Baseline Self-Efficacy × Favorability & -0.631 & 0.245 & 99.840 & -2.581 & 0.011 \\
\hline
\end{tabular}}
\end{table}

The main effect of the practice condition (tutor vs. paper) on learning gain was statistically significant, $t(62.30) = 2.38, p = .020$, indicating that students in the tutor-based condition showed greater learning gains than those in the paper-based condition. Learning gain was also significantly associated with students’ baseline self-efficacy, $t(98.40) = -2.14, p = .035$, showing that students with lower initial self-efficacy showed greater improvement. Favorability was not a significant predictor of learning gain.

\begin{figure}
  \centering
  \includegraphics[
  width=0.85\textwidth
  ]{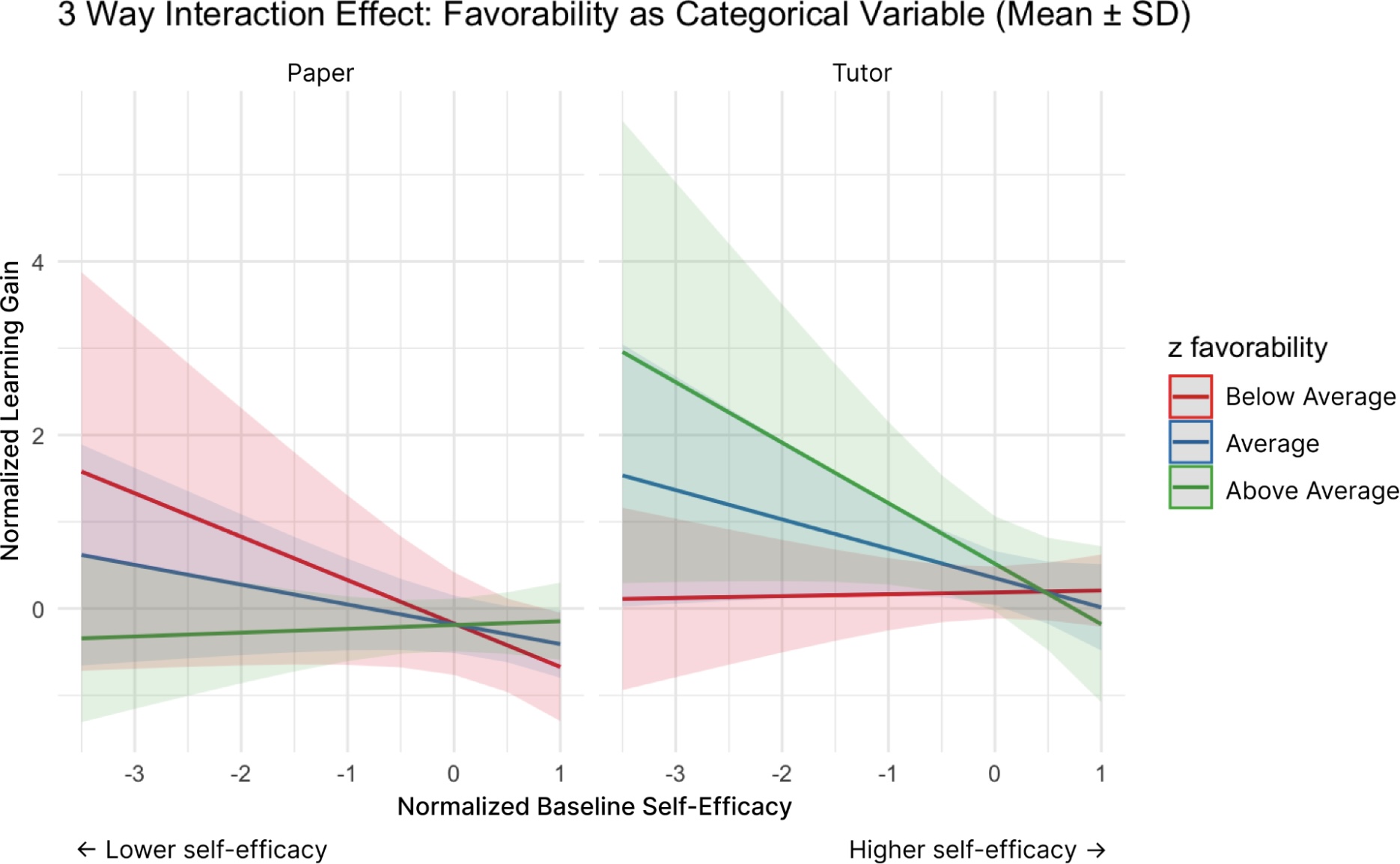}
  \caption{The three-way interaction effect on learning outcome.}
  \label{fig:3interaction}
\end{figure}

The three-way interaction between self-efficacy, favorability, and training condition was also significant, $t(99.8) = -2.58, p = .011$, indicating that the impact of initial self-efficacy on learning gains varied based on both favorability and training condition. Fig.~\ref{fig:3interaction} illustrates the three-way interaction dividing favorability into three levels for better visualization—low, medium, and high—by using the mean ± standard deviation: students below one standard deviation from the mean were classified as low favorability, those within one standard deviation as medium favorability, and those above one standard deviation as high favorability. The paper and tutor panels show learning gains with respect to the corresponding practice format (paper-based or tutor-based), respectively. The range of normalized baseline self-efficacy shown in the figure reflects the observed distribution in the sample, with relatively few values above one standard deviation. Since favorability was measured separately for each instructional condition, a student could be highly favorable toward both tutor-based and paper-based formats. As can be seen in Fig.~\ref{fig:3interaction}, for students with low baseline self-efficacy, the relative benefit of the instruction condition was influenced by how favorable they were toward that condition. Specifically, low baseline self-efficacy students who were less favorable toward paper-based practice benefited more from paper-based instruction, compared to low baseline self-efficacy students who were more favorable toward paper-based practice.  Furthermore, low baseline self-efficacy students who were more favorable toward computer-based practice benefited more from the tutor, compared to low baseline self-efficacy students who were less favorable toward computer-based practice. These findings indicate that the effectiveness of tutor-based vs. paper-based training is moderated by both self-efficacy and favorability. 

Because favorability was measured after the posttest, it is possible that the posttest results influenced the favorability results (and not the other way around). To test for this, we conducted a mediation analysis, investigating the effect of condition on favorability with potential mediation of posttest scores. We did not find any evidence that the favorability scores were associated with posttest scores, $t(51) = 1.67, p = .101$, suggesting that the favorability scores are likely to measure a general favorability toward the practice method.

\subsection{RQ2: Does an ITS Enhance Self-Efficacy?}

Table~\ref{tab:rq2} below revealed that baseline self-efficacy had a significant positive effect on post-training self-efficacy for the specific math skills practiced in the study, $t(91.94) = 5.91, p < .001$, suggesting that students who started with greater confidence in their abilities were more likely to maintain or improve their self-efficacy after training. In contrast, training conditions (tutor vs. paper), $t(57.96) = 0.49, p = .629$, and favorability, $t(90.00) = 0.96, p = .341$, did not independently predict changes in self-efficacy. We also examined interaction effects. However, none of the interactions reached statistical significance. These results indicate that, contrary to our expectation, ITS-based training did not lead to higher post-training self-efficacy, as initial self-efficacy remained the primary determinant regardless of training condition or favorability.

\begin{table}
\centering
\caption{Fixed-effects model results for RQ2.}
\label{tab:rq2}
\resizebox{\textwidth}{!}{
\begin{tabular}{lccccc}
\hline
Effect & Estimate & Std. Error & df & $t$ value & $p$ \\
\hline
Intercept & -0.026 & 0.104 & 64.525 & -0.250 & 0.803 \\
Practice Format (Paper vs. Tutor) & 0.085 & 0.174 & 57.959 & 0.485 & 0.629 \\
Baseline Self-Efficacy & 0.642 & 0.109 & 91.944 & 5.911 & 0.000 \\
Favorability & 0.088 & 0.092 & 80.995 & 0.959 & 0.341 \\
Practice Format × Baseline Self-Efficacy & 0.363 & 0.220 & 97.132 & 1.649 & 0.102 \\
Practice Format × Favorability & -0.100 & 0.195 & 99.997 & -0.513 & 0.609 \\
Baseline Self-Efficacy × Favorability & 0.136 & 0.101 & 94.779 & 1.347 & 0.181 \\
Practice Format × Baseline Self-Efficacy × Favorability & 0.108 & 0.202 & 96.467 & 0.533 & 0.595 \\
\hline
\end{tabular}}
\end{table}

\section{Discussion and Conclusion}

This study examined the role of self-efficacy and favorability in shaping learning outcomes and motivational change across tutor-based and paper-based practice. Our findings build on prior results showing that ITSs improve learning outcomes compared to paper-based practice~\cite{Borchers:2023,Borchers:2023Correction}, but go one step further by demonstrating important interactions with baseline self-efficacy and students' favorability toward the mode of instruction. In particular, the relationship between baseline self-efficacy and learning gains varied as a function of both instructional format and students’ favorability toward the instructional format, suggesting that whether students show greater learning gains from tutor-based versus paper-based practice is influenced by learners’ motivational profiles. We found no evidence that the instructional format influenced changes in self-efficacy for the specific math skills practiced in the study. Although tutor-based practice might be expected to improve students’ self-efficacy through perceived success and performance feedback that contribute to mastery experiences, a key source of self-efficacy~\cite{Bandura:1997}, the short duration of the intervention and the limited empirical literature on self-efficacy in ITS contexts may help explain why such changes did not emerge.

These findings offer a more nuanced view of prior claims about the benefits of ITSs. Prior work has shown that scaffolded support and adaptive feedback in tutor-based instruction can improve learning outcomes~\cite{Ma:2014,VanLehn:2011,Aleven:2002}, and that motivational factors like self-efficacy and favorability shape how students engage with ITSs, for example, by influencing persistence, responsiveness to feedback, and use of available supports~\cite{Bernacki:2014,McQuiggan:2007}. Our findings align with this pattern and extend prior work comparing tutor- and paper-based practice by showing that the relative benefits of these formats are not uniform, but vary systematically with learners’ motivational profiles. Among students with lower baseline self-efficacy, those who were more favorable toward the tutor showed greater learning gains during tutor practice than students who were less favorable toward it, possibly reflecting greater engagement with the tutor and its scaffolded support and timely feedback. In contrast, students who were less favorable toward the paper-based format actually improved more under the paper condition, possibly because students may have sought additional support elsewhere (e.g., from their teachers or peers) when instructional guidance was limited, although we did not directly measure such support-seeking. Alternatively, these students may have exerted greater effort when working with it, which could also account for their higher gains. Meanwhile, students with higher baseline self-efficacy showed smaller gains overall across both instructional formats, regardless of their favorability, possibly due to an association between self-efficacy and prior knowledge. Indeed, exploratory analyses showed weak but non-significant positive associations between baseline self-efficacy and pretest performance within each unit. Overall, these findings suggest that the effectiveness of instructional formats depends on how well they align with students' motivational profiles, particularly for students with low initial self-efficacy. 

While this study offers initial insights into how self-efficacy for the math topic being studied and favorability of instructional conditions influence the effectiveness of these conditions, several limitations point to directions for future research. First, due to missing performance and survey data, the final analytic sample included only 54 of the 97 participants (55.67\%), which is relatively small. This limited our statistical power and may have made it harder to detect more subtle effects or interactions. Future work should aim to replicate this study with a larger and more diverse sample. Second, favorability was measured after the posttest. Although our mediation analysis found no significant link between learning gains and favorability, suggesting that the measure may still reflect students’ initial attitudes, collecting this data earlier, ideally before the pretest, would be more informative, but would also require students to report expectations about the tutor before experiencing it. Third, the short duration of the study (five days) may have limited our ability to observe meaningful changes in self-efficacy, a relatively stable psychological construct that typically requires more sustained experience or feedback to shift appreciably~\cite{Bernacki:2014}. For future research, it would be valuable to extend the intervention over a longer timespan to better capture potential changes in students’ self-efficacy. Finally, this study focused on a single domain (linear algebra) and a specific intelligent tutoring system. As such, the extent to which these findings generalize to other subject areas or different ITS designs remains an open question. Future work should examine whether similar patterns emerge across diverse domains and learning technologies.

In sum, this study highlights the importance of considering individual motivational differences, such as self-efficacy and favorability, when evaluating the effectiveness of instructional formats. Our findings suggest that the relative benefits of tutor versus paper-based practice depend on how well the format aligns with students’ motivational profiles, specifically, a combination of their baseline self-efficacy and their preferences, consistent with prior findings suggesting that learner preferences alone have limited predictive value for achievement when considered in isolation~\cite{Schnackenberg:1998,Makransky:2020}. These insights contribute to a more nuanced understanding of personalized learning and point toward the potential value of adapting instruction to better meet the needs of diverse learners. More broadly, they lay the groundwork for future research on how instructional formats can be aligned more effectively with learners' motivational needs.

\section*{Acknowledgments}
This research was funded by the Institute of Education Sciences (IES) of the U.S. Department of Education (Award \#R305A220386).

\end{document}